\documentclass[prl,twocolumn,amsmath,amssymb,groupedaddress,superscriptaddress]{revtex4}
\usepackage{graphics}
\usepackage{epstopdf}

\def\>{\rangle}
\def\<{\langle}

\def\be{\begin{equation}}

\def\ee{\end{equation}}

\def\ttot{\textmd{tot}}

\def\ttot{\textmd{tot}}

\def\ttrans{\textmd{trans}}

\def\be{\begin{equation}}
\def\ee{\end{equation}}
\def\bea{\begin{eqnarray}}
\def\eea{\end{eqnarray}}
\usepackage{bm}
\usepackage{graphicx}

\begin{document}

\bibliographystyle{unsrt}
\title[Non-Markovian quantum transport in photosynthetic systems]{Non-Markovian stochastic description of quantum transport in photosynthetic systems}

\author{In{\'e}s de Vega}
\email{ines.devega@uni-ulm.de}
\affiliation{Max-Planck-Institut f\"ur Quantenoptik, Hans-Kopfermann-Str. 1, Garching, D-85748, Germany.}
\affiliation{Institut f\"ur Theoretische Physik, Albert-Einstein-Allee 11, Universit{\"a}t Ulm, D-89069 Ulm, Germany.}






\begin{abstract}

We analyze several aspects of the transport dynamics in the LH1-RC core of purple bacteria, which consists basically in a ring of antenna molecules that transport the energy into a target molecule, the reaction center, placed in the center of the ring. We show that the periodicity of the system plays an important role to explain the relevance of the initial state in the transport efficiency. This picture is modified, and the transport enhanced for any initial state, when considering that molecules have different energies, and when including their interaction with the environment. We study this last situation by using stochastic Schr{\"o}dinger equations, both for Markovian and non-Markovian type of interactions.

\end{abstract}


\maketitle




\section{Introduction}

Photosynthetic complexes are formed by ensembles of molecules that capture the energy from the sun, and through electron hopping transfer it to the so-called reaction center (RC): a particular part of the complex in which this energy is absorbed and used to produce chemical reactions \cite{molecularmech}.

Recent experiments have analyzed the so-called Fenna-Matthew-Olson (FMO) complex of the Green-Sulphur bacteria at cryogenic temperatures \cite{flemming07,engel07} showing that, on the one hand, the transfer process between molecules occurs in a very short time scale (of the order of several hundreds of femtoseconds), and on the other hand, quantum coherence persist in the system during most part of the process. This resuls have been extended recently in  \cite{engel10}, where it has been found evidence that quantum coherence survives in FMO even at physiological temperature for at least 300 fs. This has motivated a number of theoretical works that consider the photosynthetic complex as a quantum system, and try to analyse the basic mechanisms that explain the phenomena observed in the experiments.

Most of these works analyse in particular the FMO complex \cite{pleniohuelga08,pleniohuelga09_1,pleniohuelga09_2,rebentrost09}, and the $LH1-RC$ and $LH2$ complexes of the purple bacteria \cite{jang04,cheng06,olaya08,olaya09}. In both cases, molecules are subject to an environment of vibrating proteins, that can be described as a phonon bath. Since dissipation occurs in a much longer time scale, the bath produces basically dephasing \cite{flemming07,ishizaki09}, a mechanism that gives rise to decoherence in the system and therefore leads to a recovering of the classical behavior.
The question of how this dephasing affects the energy transport, and more generally the interplay between quantum coherence and dephasing in photosynthetic complexes has been widely discussed in the literature \cite{pleniohuelga08,pleniohuelga09_1,rebentrost09,rebentrost09_2,rebentrost08,rebentrost10}. While as described above, some experimental results suggest that quantum coherences are present in the process, their role is yet to be determined. In some situations, like quantum walks in binary tree structures  (see for instance \cite{rebentrost09} and references therein), quantum coherences might produce an exponential speed-up with respect to the classical counterpart. However, when quantum coherences are too strong, they might give rise to a decrease of the quantum efficiency, i.e. the ability to transport the energy to the reaction center. Indeed, if the system is isolated from the environment and has uniform molecular energies, quantum destructive interferences may leave the system trapped in the so-called invariant subspaces \cite{pleniohuelga08}. Once there, it cannot scape and evolve into the target state in which all the energy is transferred to the reaction center. In this scenario, a certain quantity of environment noise, or alternatively some energy mismatches in the molecules, produces enough decoherence as to drive the system out of the invariant subspaces, leading to an increase of the transport efficiency \cite{pleniohuelga08,rebentrost09,pleniohuelga09_1}. 
Environment assisted quantum transport in the FMO complex, as well as the entanglement dynamics during the process have been recently analyzed for Markovian \cite{pleniohuelga09_2,flemming09} and for non-Markovian interactions \cite{pleniohuelga09_2}.

In the present work we give a step further into these ideas, and by taking into account its symmetry properties, we analyze the transport dynamics of the LH1-RC complex both for Markovian and non-Markovian couplings with the environment. Indeed, the LH1-RC core is a highly symmetric structure, as it was revealed by pioneering X-ray crystallography experiments (see \cite{hu02} for a review). It consists in a ring of several antenna molecules coupled to a reaction center placed in the center of the ring \cite{hu97,ritz00,olaya08,olaya09}. 
The periodic arrangement of antenna molecules allows us to describe their operators and their basis in the momentum space, a representation in which several interesting features can be observed. Particularly, we consider a simple case in which all the antenna molecules are coupled uniformly to a single reaction center. In this situation, it is shown that when the antenna molecules have uniform energies and are isolated from the environment, only the zero momentum component of the initial state is transferred into the reaction center. This component corresponds precisely to the symmetric part of the initial state, what explains qualitatively and confirms some of the results obtained in \cite{olaya08,olaya09} for the same system. In this scenario, the non-zero momentum components belong to an invariant subspace. However, when the molecular energies are non-uniform and/or the coupling with the environment is considered, it is shown that the energy contained in other momentum components can also be transferred to the reaction center, which leads to an improvement of the transport efficiency. 
We show also how some of these results can be extended to the case of several reaction center molecules non-uniformly coupled to the antenna molecules.

The effects of the environment on the the transport dynamics is described by considering a quantum open system (QOS) approach. Within this framework, the dynamics of the system is usually described with a master equation for its reduced density operator (see \cite{breuerbook} for a review), obtained after tracing out the environment degrees of freedom. An alternative to this scheme are the so-called stochastic Schr{\"o}dinger equations (SSEs). They evolve system vectors with the property that the average over their projector is equal to the density matrix. Some SSEs describe a deterministic evolution randomly interrupted by discontinuous quantum jumps, as described in \cite{plenioknight98} for the Markov case and in \cite{piilo08} for the non-Markovian case. The late equations have recently been applied to analyze the dynamics of the FMO complex \cite{rebentrost10}. A second type of stochastic equations lead to trajectories where the noise acts continuously at each time step of the evolution. This so-called quantum state diffusion scheme has been derived for both Markovian and non-Markovian interactions \cite{YDGS99,SDG99,strunz01,devega05,devega06}, and it will be used here to describe some of the features of the LH1-RC complex.

The stochastic scheme presents several advantages with respect to the use of the master equations: first, this formalism allows us to develop an approximation, the so-called post-Markov approximation, alternative to commonly used weak coupling approximation and valid for different parameter regimes. While the weak coupling approximation is valid for  $\Gamma_{\textmd deph}\ll\gamma$, where $1/\gamma$ is the relaxation time of the environment, and $1/\Gamma_{\textmd deph}$ is the system dephasing time, the post-Markov approximation is valid for weakly coupled systems, such that the mean coupling $J$ is such that $J\ll\gamma$. Second, stochastic Schr{\"o}dinger equations evolve the system state, that has say $M$ components ($M$ being the dimension of the system Hilbert space), while master equations evolve a matrix of $M\times M$ elements. Considering a single excitation in the system (an approximation that will be later explained in the context of photosynthetic systems), $M$ is just the number of component particles or molecules. Hence, for a relatively large system like the LH1-RC complex, and provided that the number of stochastic trajectories needed to obtain the relevant quantities is not too large, SSEs might represent an advantage from the computational point of view. Note that, for instance, for $M=33$ molecules and within the one excitation sector, a master equation deals with a Liouvillian of dimension $33^4\sim 10^6$, which is already very close to the computational limit.

The paper is organized as follows. First we analyze the system in the momentum space. Within this framework, the relevance of the initial states in the transfer efficiency is discussed. Then it is shown how the static disorder in the energies of the antenna molecules, as well as their coupling to the environment can increase this efficiency. In order to describe the system coupled to the environment, SSEs are presented together with the post-Markov approximation. In that way the transport efficiency of the system undergoing dephasing is analyzed both for Markovian and non-Markovian interactions.

\section{A simple model to study LH1-RC}

Let us now analyze the energy transport in the LH1-RC core of purple bacteria, based in a simplified picture of the more complex model described in \cite{olaya08,olaya09} and depicted schematically in Figure (\ref{intu}). 
For simplicity, the antenna molecules of the ring, as well as the reaction center, are considered as two-level systems. A more realistic situation in which the reaction center is modeled by a molecular pair is found in \cite{olaya08}. However, we expect that our simplified model captures some of the main aspects of the quantum transport phenomena occurring in the system. On the other hand, the rate at which purple molecules absorb photons (approx. one photon every two hours) is much smaller than the rate at which the energy is transfered to the reaction center. Thus, it is reasonable to assume that only one excitation is present in the system during the energy transport process.

In this situation, the state of the system molecules can be described in terms of a basis $\{|1\rangle_j\}$ for $j=1,\cdots,M+1$, where $|1\rangle_j=|0\cdots 1_j \cdots 0\rangle$ 
is the state where only the $j$th molecule is in its excited electronic state 
. Note that antenna molecules are labeled with an index running from $1$ to $M$, while the reaction center is labeled with the index $M+1$.  In that way, the effective system Hamiltonian can be written as
\bea
H_S&=&\sum_{j=1}^M \omega_j \sigma_j^+\sigma^-_j+\sum_{i,j=1}^MJ_{ij}\sigma_j^+ \sigma^-_j+\sum_j \Gamma_j (\sigma_j^+\sigma^-_{M+1}\nonumber\\&+&h.c.)+(\omega_{M+1}+i\kappa)\sigma_{M+1}^+\sigma^-_{M+1}.
\label{space}
\eea
Here, $\omega_i$ are the energies of the antenna molecules, while $J_{ij}=J_{ji}$ are the electronic rates that describe the electron hopping from site $i$ to site $j$. Also, $\sigma^-_j=|0\rangle_j\langle 1|$ (similarly $\sigma_j^+$) is the spin ladder operator describing the transition from one to zero (zero to one) excitations at site $j$, and $\sigma^-_{M+1}=|0\rangle_{M+1}\langle 1|$ (similarly $\sigma_{M+1}^+$) is describing the transition between one and zero (zero to one) excitations in the reaction center. The non-Hermitian term of the former Hamiltonian $i\kappa\sigma_{M+1}^+\sigma_{M+1}$, has been added to phenomenologically describe the energy sink placed in the reaction center, which in addition has an energy $\omega_{M+1}$. The coupling coefficients of each molecule $j$ with respect to the RC molecule are designed as $\Gamma_j$. 

As noted above, we consider here that there is a single molecule in the RC. However, our discussion can be easily extended to the case of having $D>1$ molecules in the RC \cite{olaya08,hu97}. In that situation, we should consider that the interaction Hamiltonian between each of the $j$ antenna molecules and the each of the $l$ molecules in the RC can be written as $H_{LH1-RC}=\sum_{l=1,D}\sum_{j=1,M} \Gamma^l_j (\sigma_j^+\sigma^-_{l,RC}+h.c.)$, where $\sigma^-_{l,RC}$ design the spin ladder operators corresponding to the RC molecules. This case will be analyzed in detail elsewhere.

Let us now represent the Hamiltonian (\ref{space}) in the momentum space. To this order, we consider the Fourier decomposition of spin operators for the antenna molecules,
\begin{eqnarray}
\sigma^-_{ j} =\frac{1}{\sqrt{M}}\sum_{q} e^{iq r_j}\sigma^-_q
\label{TF}
\eea
for $j=1,\cdots M$. Here, $r_j=d_0 j$, and $d_0$ the distance between different molecules in the ring. A similar expression exists to relate the position and momentum basis, $|1\rangle_j=\frac{1}{\sqrt{M}}\sum_q e^{iqr_j}|q\rangle$.

Having a system with periodic boundary conditions allows us to derive a crucial property, that may lead to important simplifications when expressing the system Hamiltonian in the momentum space. For a one dimensional periodic system like the molecules in the ring, we know that $\sigma^-_j=\sigma^-_{j+M} $. According to (\ref{TF}), this means that $e^{i q r_M}=1$, what implies that momentum $q$ can only take discrete values $q_m=\frac{2 \pi m }{d_0 M}$, with $m=1,\cdots,M$. This leads to the equation
\begin{eqnarray}
\frac{1}{M}\sum_{j}^M e^{i (q_m-q_n) j d_0 }=\delta_{m,n}.
\label{delta}
\end{eqnarray}
Using this property, together with the definition (\ref{TF}) and a similar one for $\sigma_j^+$, a simple form for $H_S$ in the momentum space can be obtained,
\bea
H_S&=&\sum_{q,q'}\omega_{q,q'}\sigma_q^\dagger\sigma^-_{q'}+\sum_q J_q \sigma_q^\dagger \sigma^-_q+\sum_q \Gamma_q (\sigma^+_q\sigma^-_{M+1}\nonumber\\&+&h.c.)
+(\omega_{M+1}+i\kappa)\sigma_{M+1}^+\sigma^-_{M+1}.
\label{momentum}
\eea
with $\omega_{q,q'}=\sum_je^{i(q-q')}\omega_j$,
\bea
\sigma_q=\frac{1}{\sqrt{N}}\sum_{j=1}^M \sigma_j e^{-i q r_j},
\eea
$J_q=\sum_j J_j e^{-i q r_j}$, and $\Gamma_q=\frac{1}{M}\sum_j \Gamma_j e^{-i q r_j}$.
Depending on the profile of $\Gamma_q$, different momentum $q$ are directly coupled to the RC. 

Let us assume an idealistic situation where all the dipole moments of the molecules within the system are approximately aligned. In that situation, we can consider $\Gamma_i\approx \Gamma$, so that  $\Gamma_q\approx \Gamma \delta_{q0}$ so that according to (\ref{momentum}), only the zero momentum spin-wave, $\sigma_0$ couples to the reaction center. This case corresponds to the configuration considered in the first part of the paper \cite{olaya08}, and is valid when the mean value of the set of $\Gamma_i$, is much smaller than their dispersion.

\begin{figure}[ht]
\centerline{\includegraphics[width=0.45\textwidth]{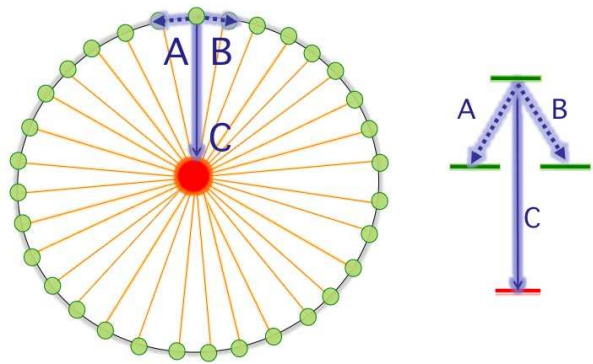}}
\caption{Schematic representation of the LH1-RC complex, with $32$ antenna molecules distributed along a ring around the reaction center. The role of quantum coherence in the transport is intuitively explained as follows: When there is an excitation in a molecular ring with uniform energies, the quantum paths $A$ and $B$ interfere destructively with each other, so that path $C$ (and therefore the transmission to the reaction center) is favored.  Now, if we start with a symmetric state (corresponding to $q=0$) this basic process will occur simultaneously at every site, giving rise to a perfect transfer of the excitation to the reaction center. This destructive interference is similar to the one that appears in quantum binary tree structures \cite{rebentrost09}. A schematic diagram of the basic building block of this process is shown at the right hand side of the figure. \label{intu}}
\end{figure}

On the one hand, if we assume uniform molecular energies $\omega_j=\omega$, then $\omega_{q,q'}=\omega\delta_{q,q'}$ and the Hamiltonian (\ref{momentum}) does not describe any transfer process between different momentum components. In that case, if we start with an initial state that in the momentum basis is written as $|\Psi(0)\rangle=\sum_q {\mathcal A}_q (0)|q\rangle$, the final transmission probability to the reaction center is $P_{\ttrans}=P_T(\infty)=|{\mathcal A}_{q=0}(0)|^2$, with  $P(t)=1-\sum^{M+1}_{j=1}\langle\sigma_j^+(t)\sigma^-_j(t)\rangle$. 

Hence, according to the picture described in \cite{pleniohuelga09_1}, the subspace spanned by the vectors $\{q\}$, with $q\neq 0$ is an invariant subspace from which the excitation cannot escape and be transfered to the reaction center. This may explain in a very intuitive way some of the results of \cite{olaya08,olaya09}, in which it is shown that symmetrical initial conditions give rise to a better efficiency. Indeed, for a highly non-symmetric initial state, as $|\Psi(0)\rangle=|1\rangle$, that in the momentum basis has coefficients ${\mathcal A}_q(0)=\frac{1}{\sqrt{M}}e^{iqr_1}$, the transmission probability is just $P_{\ttrans}=1/M$. On the contrary, for a symmetric state as $|\Psi_0\rangle=\frac{1}{\sqrt{M}}\sum_j|i\rangle$, with coefficients in the momentum basis ${\mathcal A}_q=\delta_{q0}$, the transmission probability is the maximum possible $P_T(t=\infty)=1$. Figure \ref{intu} offers a possible explanation of the role of coherences in the process.

On the other hand, if local site energies are non-uniform, the first term in the Hamiltonian (\ref{momentum}), $\sum_{q,q'}\omega_{q,q'}\sigma_q^\dagger\sigma^-_{q'}$ leads to an energy transfer between different momentum components. In that case, the $q=0$ component can eventually be \textit{repopulated} by other momentum components. This allows to inject more energy into the reaction center, and therefore to an increase in the transport efficiency. This can be observed in Fig. (\ref{DeltaOmega}), where the evolution of the transmitted energy $P_T$ is represented for uniform and non-uniform energies. 

\begin{figure}[ht]
\centerline{\includegraphics[width=0.45\textwidth]{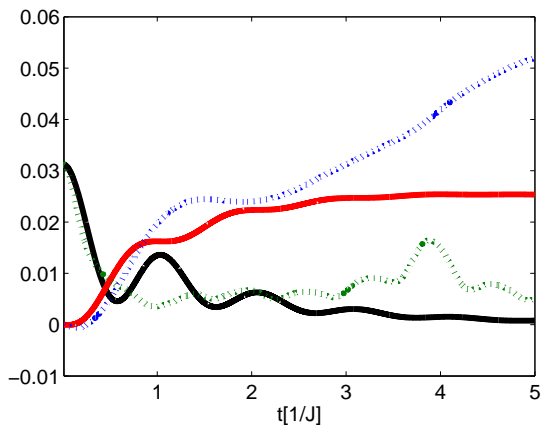}}
\caption{Evolution of the quantity $P_{q=0}(t)=\langle\sigma^+_{q=0}(t)\sigma^-_{q=0}(t)\rangle$ and transmission probability, $P_T(t)$, for two cases: A closed system in which every molecule has the same energy $\omega$ (black and red solid lines for $P_{q=0}$ and $P_T$ respectively), and a closed system in which molecular energies are randomly distributed as $\omega_j=\omega_0 \xi_j$, with $\xi_j$ a randomly distributed number between $0$ and $1$ and $\omega_0=20 J$ (green and blue dashed lines for $P_{q=0}$ and $P_T$ respectively). In the first case, $P_{q=0}(0)$ gives the final transmission probability. In the second case the population of the zero momentum spin wave suffers some revivals produced by the energy static disorder, in such a way that the transmission probability is increased.
\label{DeltaOmega}}
\end{figure}

Throughout the paper, all the quantities are given in units of the typical transfer rate between near neighbours in the photosynthetic transport, $J$. According to \cite{ishizaki09} typical values for this quantity are $J=20cm^{-1}\approx 1.667 ps$.

An alternative way to observe an increase of the transport efficiency is by realizing that molecules in the complex are not isolated, but rather they appear to be coupled to a phononic bath produced by the surrounding proteins \cite{adolphs06}. Since details about the microscopic model for this interaction are still unknown, we consider a phenomenological model Hamiltonian. As discussed in the introduction, the coupling produces mostly dephasing in the antenna molecules. Hence, our model Hamiltonian essentially accounts to modeling pure dephasing in the presence of a fluctuating field, and can be written as follows
\bea
H_j=\sum^{M}_{j=1} A_j {\mathcal B}_j,
\label{int}
\eea
where ${\mathcal A}_j= \sigma_j^+ \sigma^-_j $ and ${\mathcal B}_j=\sum_\lambda \hat{g}_{j,\lambda}(a_\lambda^\dagger+a_\lambda)$, with $a_\lambda$ ($a_\lambda^\dagger$) the annihilation (creation) operators on the environment Hilbert space, and $g_{i,\lambda}$ describe the coupling strength of the $i$ molecule to the environment. The Hamiltonian (\ref{int}) can be written in the momentum space as
\bea
H_j=\sum_{q,q'}B_{q,q'}\sigma_q^\dagger\sigma^-_{q'}
\label{momentum3}
\eea
where $B_{q,q'}=\sum_je^{i(q-q')}B_j$. The Hamiltonian (\ref{momentum3}) describes energy transfer between different momentum components of the antenna molecules, but only for inhomogeneous couplings, when the environment operators $B_j$ act differently over different molecules $i$. This means that the more uniform the interaction is, the less efficient is the mixing mechanism, what leads to a less efficient transport.

As noted above, we have considered molecular dipole moments are approximately aligned, which as analyzed in \cite{hu97,olaya08}, does not correspond to a realistic situation. However, we note that the representation in the momentum space here described, as well as some of the conclusions obtained, are useful even in the more general case in which the couplings $\Gamma_j$ (even for different reaction centers $\Gamma_j^l$) are not uniformly distributed. To be more specific, we have introduced the data available in \cite{hu97} for positions and dipole vectors of both antenna and reaction center molecules, in order to calculate the corresponding $\Gamma_j^l$. We have then computed each $\Gamma^l_q$ (as the Fourier transform of the set of $\Gamma_j^l$ corresponding to each of the $l=4$ reaction centers described in \cite{hu97}), and found that they are functions that present peaks at particular values of $q$. These peaks will correspond to the momentum values $q$ that will be preferentially transfered to the reaction center. In addition, we have seen that for other values of $q$, all the functions $\Gamma^l_q$ are comparatively very small. Again, if there is no mechanism present in the system that mixes different momentum components (i.e. non-uniformities in the molecular energies, or dephasing), this momentum values in which $\Gamma^l_q\approx 0$ will correspond to an invariant subspace: they will not be able transfer their energy to the sink. Further details of this realistic model will be analyzed elsewhere.



In the next sections, let us analyse in detail the mixing mechanism produced by a dephasing, as well as its effects in the transport.

\section{Reduced propagator: a tool to compute the dynamic of a many-body QOS}
\label{ch2sec2}

In order to describe the system dynamics, we assume a general model Hamiltonian $H_\ttot$ for a system with Hamiltonian $H_S$, coupled to an environment with Hamiltonian $H_B$,
\begin{eqnarray}
H_\ttot&=&H_S+ H_B + H_j \cr
&=& H_S + \sum_\lambda \omega_\lambda a_\lambda^\dagger a_\lambda + \sum_{j=1}^{M} \sum_\lambda g_{j,\lambda} \left(a_\lambda^\dagger L_j + L_j^\dagger a_\lambda\right),
\label{chapdos1}
\end{eqnarray}
where $L_j$ is the coupling operator corresponding to the particle $i$ and acts on the Hilbert space of the system, and the $\omega_\lambda's$ are the frequencies of the harmonic oscillators. The Hamiltonian (\ref{int}) is recaptured by considering $L_j=L^\dagger_j=\sigma_j^+\sigma^-_j$.

The total wave function of the system, described with the Hamiltonian (\ref{chapdos1}), evolves from its initial value $\mid\Psi_0 \rangle$ as,
\begin{eqnarray}
\mid\Psi_t \rangle ={\mathcal U} (t,0)\mid\Psi_0 \rangle
\label{chapdos2}
\end{eqnarray}
where ${\mathcal U} (t,0)$ is the evolution operator in interaction picture,
\begin{eqnarray}
{\mathcal U} (t,0)=e^{iH_B t}e^{-iH_{tot}t}.
\label{chapdos3}
\end{eqnarray}

We can represent the state (\ref{chapdos2}) in the basis of the environment, which is here chosen as the Bargmann coherent state basis \cite{quantumoptics}.
In terms of this basis, (\ref{chapdos2}) is written as
\begin{equation}
\mid \Psi_{t}\rangle=\int d\mu(z_1)G(z^*_1 z_0|t)\mid\psi_0\rangle\mid z_1 \rangle,
\label{psitot}
\end{equation}
with the notation $\mid z_\beta \rangle=\mid z_{\beta,1}\rangle\mid z_{\beta,2}\rangle\cdots\mid z_{\beta,\lambda}\rangle\cdots$ for the state of the environment, given by a tensor product of the states of all the $\lambda$ environmental oscillators. In our case, we are dealing with two states $z_0$ and $z_1$ corresponding respectively to $\beta=0,1$. In addition, we have defined the quantity $d\mu (z)$ as a Gaussian measure given by
\begin{eqnarray}
d\mu(z_\beta)=\frac {d^2 z_\beta }{\pi}e^{-|z_\beta|^2}=\prod_{\lambda} \frac {d^2 z_{\beta,\lambda} }{\pi}e^{-|z_{\beta,\lambda}|^2}.
\end{eqnarray}
Equation (\ref{psitot}) is obtained by introducing the closure relation for the Bargmann coherent states, $\int d\mu(z)|z\rangle\langle z|=I$ in (\ref{chapdos2}), and considering as initial state $|\Psi_0\rangle=|\psi_0\rangle|z_0\rangle$, with $|\psi_0\rangle$ the initial state of the open system, and $|z_0\rangle$ is the initial state of the environment. On the other hand, we have defined the function \cite{strunz01,devega05,devega06}
\begin{equation}
G(z^*_1 z_0 |t 0)=\langle z_1\mid {\mathcal U} (t,0) \mid z_0 \rangle,
\label{propag}
\end{equation}
that corresponds to the so-called system reduced propagator. This quantity acts on the system Hilbert space, giving the evolution of system state vectors from $0$ to $t$, conditioned that in the same time interval the environment coordinates go from $z_0$ to $z_1$.



The phonon environment surrounding the antenna molecules in a photosynthetic complex is considered to be in a thermal state, and not in a pure state as $|z_0\rangle$. Therefore, in order to describe the molecular dynamics a more general initial condition should be considered,
\begin{eqnarray}
\rho_{tot}(0)=\int d\mu(z_0) P_T(z_0 , z^*_0) |z_0 \rangle\langle z_0|\otimes|\psi_0 \rangle\langle\psi_0 |,
\label{thermal}
\end{eqnarray}
where $P_T(z_0 , z^*_0)$ is the coherent state diagonal distribution \cite{quantumoptics} corresponding to a thermal reservoir.

Any expectation value of a system operator $\mathcal{A}$ can be obtained as $\langle \mathcal{A}(t)\rangle=Tr_S(\rho_S (t)\mathcal{A})$, with $\rho_S(t)=Tr_B(\rho_{\ttot}(t))=Tr_B\large(\int d\mu(z_0) P_T(z_0 , z^*_0) {\mathcal U}(t)|z_0 \rangle\langle z_0|\otimes|\psi_0 \rangle\langle\psi_0 |{\mathcal U}^\dagger(t)\large$. Introducing twice the closure relation of the environment, we get the following expression for $\rho_s(t)$ in terms of the reduced propagators,
\begin{eqnarray}
\rho_s (t)&=&\int d\mu(z_1) \left(\int d\mu(z_0) P(z_0 , z^*_0) \right.\nonumber\\
&&\left. G(z^*_1 z_0 |t 0)|\psi_0 \rangle \langle \psi_0 | G^{-1}(z_0^* z_1 | 0 t ) \right).
\label{rhos}
\end{eqnarray}
Note that this equation is just a representation of the reduced density matrix in terms of a coherent state basis of the environment. The reduced propagators are then quantities completely determined by the knowledge of the initial and the final state of the environment. However, this initial and final states are described by an infinite set of complex quantum numbers, corresponding to each of the harmonic oscillators that compose the environment,  distributed according to the measures $d\mu(z_1)$ and $d\mu(z_0)P_T$ respectively. Therefore, the integrals appearing in (\ref{rhos}) are infinite and multidimensional, and a Monte-Carlo method should be used to solve them numerically. With this method, the integrals are sampled by choosing a random a set of coherent state coordinates $z$ that are distributed according to their measure. In that situation, the reduced propagator of the system $G(z^*_1,z_0|t, 0)$ become a stochastic object \cite{strunz96,strunz01,devega05,devega06}.

We stress that the $P$ function appearing in (\ref{rhos}) is by definition a quantum distribution, and hence it can take negative values. However, for a thermal environment it is positive definite, as it is required in order to perform the sampling over the coherent states.

From equation (\ref{rhos}), it is clear that two important ingredients are needed in order to compute system quantum mean values with this method: first, we should be able to evolve the system propagator for different initial and final states of the environment. In other words, for different noise histories. Second, we should perform the average over an ensemble of trajectories that is large enough to ensure that the Montecarlo integrals are sampled properly. With the Montecarlo method, the better the sampling is, the closer we are to the exact solution of the problem, except for the approximations made, in the equations of motion for the reduced propagators. We consider that the sampling is good enough when the result converge to a value that remains fixed even if the number of trajectories included in the sampling is increased.

\subsection{Evolution equation for the reduced propagator}

\label{ch2sec3}

The evolution of the reduced propagator can be derived following a similar procedure as in \cite{devega05,devega06}, and reads as follows (see further details of the derivation in Appendix A),
\begin{eqnarray}
&\frac{\partial G(z_1^* z_0|t0)}{\partial t}&=-i H_S+\sum_{j=1}^M \left(L_j z^{*}_{j1,t}-L_j^\dagger z_{j0,t}\right)G(z^*_1 z_0|t0)\cr&-&\sum_{p,j=1}^M L_p^\dagger
\int_{0}^{t} d\tau \alpha_{pj}(t-\tau) \cr
&\times&\langle z_1 | {\mathcal U}(t,0)L_j(\tau,0)|z_0 \rangle,
\label{chapdos13}
\end{eqnarray}
where for each particle $j$, we have defined the functions
\begin{equation}
z_{j\beta,t}=i\sum_\lambda g_{j,\lambda} z_{\beta,\lambda}e^{-i \omega_\lambda t},
\label{chapdos14}
\end{equation}
for $\beta=0,1$, and
\begin{equation}
\alpha_{pj}(t-\tau)=\sum_\lambda g_{p,\lambda} g^{*}_{j,\lambda }e^{-i \omega_\lambda (t-\tau)}.
\label{chapdos15}
\end{equation}
The quantity $\alpha_{pj}(t-s)$ is the time autocorrelation function of the noise $z_{j1,t}$, as it can be easily verified by computing the average ${\mathcal M}[z_{p1t}z^{*}_{j1\tau}]$ with respect to the measure $d\mu(z_1)$. From equation (\ref{chapdos13}) it can also be seen why the correlation function $\alpha_{pj}$ is responsible of the dependence of the evolution of the system over its past history. Particularly, it is the kernel of an integral that goes from the initial time $0$ to the actual time $t$. In that situation, the slower the correlation function decays (i.e. the larger is the correlation time $\tau_c$), the more contributions appear from past times.

In order to have a closed evolution equation for the reduced propagator, the matrix element $\langle z_1 | {{\cal U}}(t,0)L_j(\tau,0)|z_0 \rangle$ appearing in the last term of the equation (\ref{chapdos13}) should be expressed in terms of $G(z_1^*z_0|t0)$. In order to do so, some approximation need to be considered, such that the matrix element can be written as a certain system operator $O(z_0 z^*_{1,i},t,\tau)$ multiplied by the propagator,
\begin{eqnarray}
\langle z | {\mathcal U}(t,0)L_j (\tau,0)|z_0 \rangle&=& \langle z |L_j(\tau,t_{i}) {\mathcal U}(t,0)|z_0 \rangle \nonumber\\
&\approx&O_j(\tau,t,z_1^* z_0)G(z^*_1 z_0|t 0).
\label{chapdos18}
\end{eqnarray}
Inserting (\ref{chapdos18}) in (\ref{chapdos13}), we get the following closed evolution equation for the general reduced propagator,
\begin{eqnarray}
\frac{\partial G(z^*_1 z_0|t0)}{\partial t}&=&\big(-i H_S+\sum_{j=1}^M (L_j z^{*}_{j1,t}\cr
&-& L_j^\dagger z_{j0,t})\big)G(z^*_1 z_0|t0)-{\mathcal T}
\label{chapdos19}
\end{eqnarray}
where
\bea
{\mathcal T}=\sum_{p,j=1}^ML_p^\dagger \int_{0}^{t} d\tau\alpha_{pj}(t-\tau) O_j(\tau,t,z_1^* z_0)G(z^*_1 z_0|t 0).
\eea
Once an approximate form of $O_j(\tau,t,z_1^*z_0)$ is known, the equation (\ref{chapdos19}) can be used to integrate the reduced propagator along with its initial conditions $G(z^*_1 z_0|t t)=\exp{(z^*_1 z_0)}$.

In the former equations, the operator $O_j$ is constructed with the post-Markov approximation, first introduced in \cite{YDGS99} for a single particle quantum open system. This approximation is presented in more detail in the next section.

\section{Post-Markov approximation}
\label{ch2sec4}
We consider the post-Markov approximation in order to calculate a particular expression for $O_j(\tau,t,z_1^*z_0)$ in (\ref{chapdos18}).
To this purpose, we write the last term of  (\ref{chapdos13}) as
\begin{eqnarray}
{\mathcal T}&=&\sum_{p,j=1}^M\int_0^t d\tau \alpha_{pj}(t-\tau)L_p^\dagger\langle z_1 |L_j(\tau,t) {\mathcal U}(t,0)|z_0 \rangle,
\label{chapdos33}
\end{eqnarray}
and perform an expansion of $L_j(\tau,t)=L_j(\tau-t )$ in Taylor series $(t -\tau)$,
\begin{eqnarray}
&&L_j(\tau-t )=L_j +\left. \frac{dL_j(\tau-t )}{dt }\right|_{\tau=t}(\tau-t)\nonumber\\
&+&\left.\frac{d^2 L_j(\tau-t )}{dt^2 }\right|_{\tau=t}(\tau-t)^2 +{\mathcal O}((\tau-t)^3 ).
\label{eleideriva}
\end{eqnarray}

This expression, when inserted in (\ref{chapdos33}), gives rise to a series expansion of $T$ of the form ${\mathcal T}=\sum_n {\mathcal T}_n$ for $n=0,1,2\cdots$, with
\begin{eqnarray}
{\mathcal T}_n&=&\sum_{p,j=1}^M\int_0^t d\tau \alpha_{pj}(t-\tau)L_p^\dagger O^n_j (\tau,t, z_1^*z_0) G(z^*_1 z_0|t0)
\label{series}
\end{eqnarray}
where
\bea
&&O^n_j (\tau,t|z_1^*z_0)G(z^*_1 z_0|t0)=(\tau-t)^n
\nonumber\\
&& \langle z_1|\left.\frac{dL^n_j(\tau-t )}{dt^n }\right|_{\tau=t}{\mathcal U}(t,0)|z_0\rangle,
\eea
In the notation of (\ref{chapdos18}), we can write $O_j(\tau,t,z_1^*z_0)=\sum_n O^n_j (\tau,t, z_1^*z_0)$.
Note that inside the time integrals appearing in (\ref{series}), the maximum value for the quantity $(\tau-t)^n$ is the decaying time of the correlation function $\alpha_{pj}(t)$, or correlation time $\tau_c$. Hence, the $n$-th term of the series is of order $\tau_c^n$.

The terms $n=0,1$ of the series for our system, described by the Hamiltonian (\ref{space}), can be written as
\begin{eqnarray}
{\mathcal T}_0=\sum_{p,j=1}^M\int_0^t d\tau \alpha_{pj}(t-\tau)L_p^\dagger L_j G(z^*_1 z_0|t0),
\label{series0a}
\end{eqnarray}
with $L_j=\sigma_j^+\sigma_j$, and
\begin{eqnarray}
{\mathcal T}_1&=&
i\sum_{pj=1}^M\int_0^t d\tau \tau {\mathcal R}_{pj}(\tau) G(z^*_1 z_0|t0)
\label{series2a}
\end{eqnarray}
where
\bea
{\mathcal R}_{pj}(\tau)=\alpha_{pj}(\tau)L_p^\dagger\sum_l\bigg(J_{jl} \sigma^+_j\sigma_l
-J_{lj}\sigma_l^+\sigma_j\bigg).
\eea

Within the single excitation sector and considering $L_p=\sigma_p^\dagger\sigma_p$, the former equation can be further simplified, since $L_p^\dagger \sigma^+_j\sigma_l=\sigma_j^+\sigma_l\delta_{pj}$. From (\ref{series0a}) and (\ref{series2a}) we can infer that $O^0_j(\tau t|z_1^* z_0)\equiv O^0_j(\tau t)=L_j$, and $O^1_j(\tau t|z_1^* z_0)\equiv O^1_j(\tau t)=\sum_l(J_{jl}\sigma^+_j\sigma_l
-J_{lj}\sigma_l^+\sigma_j)$. Hence, for $n=0,1$ the operator $O_j (\tau,t|z_1^*z_0)$ does not depend on the noises, and can be written as $O_j(\tau, t)=\sum_n O^n_j(\tau,t)$. Further details of the calculation are explained in Appendix B.

In order to approximate ${\mathcal T}$ with a series expansion, it is important to establish the conditions under which this series converge, so that we can just keep the lower orders. This issue discussed in Appendix C.

\section{Master and stochastic equation for $L_j=L^\dagger_j$}

Stochastic equations can be used to derive the corresponding master equation \cite{YDGS99,devega06}. This is done by considering the derivative of $\rho_s$ according to the definition (\ref{rhos}), then inserting the evolution equation for the reduced propagator (\ref{chapdos13}), and finally performing analytically the averages over the environment degrees of freedom. In that way, and for the simple case of an hermitian coupling $L_j=L^\dagger_j=A_j$, the master equation up to first oder in $n$ can be written as
\bea
\frac{d\rho_s}{dt}&=&i[H_S,\rho_s]+\sum_{p,j=1}^M\int_0^t d\tau \alpha_{pj}^{T*} (t-\tau)[A_j,\rho_s O_{j}^\dagger(t,\tau)]\nonumber\\
&+&\sum_{p,j=1}^M\int_0^t d\tau \alpha^T_{pj} (t-\tau)[O_j(t,\tau)\rho_s, A_j],
\label{master}
\eea
with $\alpha^T_{pj}(t)=\sum_\lambda g_{j\lambda}g_{p\lambda} (\cosh(\omega_\lambda \beta)\cos(\omega_\lambda t)-i\sin(\omega_\lambda t))$, and $O_j(\tau,t)=\sum_{n=0,1}O^n_j(\tau,t)$ derived in the former section. Here, the constant $\beta=1/(K_B T)$, with $K_B$ the Boltzmann constant, and $T$ the temperature of the reservoir. 

With some calculations, it can be seen that the master equation (\ref{master}), valid for an hermitian coupling, can be obtained also as $\frac{d\rho_s(t)}{dt}=\frac{d}{dt}(\int d\mu(z)G(z^* 0|t0)|\psi_0\rangle\langle\psi_0|G^\dagger(z^*0|t0))$, by considering the following stochastic evolution,
\begin{eqnarray}
\frac{\partial G(z^* 0|t0)}{\partial t}&=&\big(-i H_S+\sum_{j=1}^M A_j z^{*}_{jt}\big)G(z^* 0|t0)-{\mathcal T}
\label{chapdos192}
\end{eqnarray}
with
\bea
{\mathcal T}\approx\sum_{p,j=1}^MA_j \int_{0}^{t} d\tau\alpha^T_{pj}(t-\tau) O_j(t,\tau)G(z^* 0|t 0).
\eea
Note that this stochastic evolution is much simpler than (\ref{chapdos19}), since it depends on a single noise such that $\int d\mu(z)z_t^* z_\tau=\alpha^T(t-\tau)$. For that reason, this will be the stochastic equation to use in this work.

\subsection{Evolution equations in the system basis}
\label{evol}

We now write the stochastic (\ref{chapdos192}) and the master equation (\ref{master}) in the system basis corresponding to the Hamiltonian (\ref{space}).

In this basis, the evolution of the wave vector $|\psi_t(z)\rangle=\sum_j a_j(t)|1\rangle_j$, where the coefficients $a_j(t)=\langle 1|_jG(z^* 0|t0)|\psi_0\rangle$ represent the probability amplitude that the $i$ molecule is excited, conditioned to an initial and final state of the environment $0$ and $z$ respectively. For simplicity, we have excluded in the notation the dependency of $a_j(t)$ over these environment states. 

Considering the evolution equation for the propagator $G(z^*_1z_0|t0)$ given by (\ref{chapdos19}), with the last term approximated with a post-Markov expansion up to second first order, ${\mathcal T}=\sum_{n=0,1} {\mathcal T}_n$, it is found that the evolution equation of a single coefficient $a_j(t)$ for $j=1,\cdots M$ is
\bea
\frac{da_j(t)}{dt}&=&-i\bigg(\sum_{p=1}^M J_{jp} a_p(t)+\omega_j a_j(t)\bigg)
-i \Gamma a_{M+1}\nonumber\\&+&z^{*}_{jt} a_j(t)
-{\mathcal O}^0_{jj}a_j(t)\cr
&-&i\sum_{p=1}^M\bigg({\mathcal O}_{jj}^1 J_{jp}a_p(t)-{\mathcal O}_{jp}^1 J_{pj}a_j(t)\bigg),
\label{stoch}
\eea
and for the reaction center labeled as $M+1$,
\bea
\frac{da_{M+1}(t)}{dt}&=&-i\bigg(\omega_{M+1}-i\kappa\bigg) a_{M+1}(t)\cr&-&i \Gamma \sum_{j=1}^M a_j(t).
\label{stoch2}
\eea
Here, we have defined
\bea
{\mathcal O}^n_{pj}=\int_0^t d\tau (t-\tau)^n \alpha^T_{pj}(t-\tau).
\eea
In this paper, we will consider the same correlation function as in \cite{ishizaki09}, defined as
\begin{eqnarray}
\alpha^T_{pj}=\delta_{pj}g_j(\frac{2}{\beta}+i 2 \gamma)e^{-\gamma t}
\label{correl}
\end{eqnarray}
where local noise and high temperature conditions, $\beta\hbar \gamma=0.25<1$, have been assumed. This correlation function corresponds to an environment with a Drude-Lorentz spectral density
\begin{eqnarray}
J_j(\omega)=2g_j \frac{\omega\gamma_j}{\omega^2+\gamma_j^2}.
\end{eqnarray}
From equation (\ref{master}), the corresponding master equation can be written as
\bea
\frac{\rho_{pj}}{dt}&=&i\sum_{l=1}^M\bigg(J_{pl}(1-\delta_{p,M+1})\rho_{lj}-J_{lj}(1-\delta_{j,M+1})\rho_{pl}\bigg)\nonumber\\
&-&i(\omega_p-\omega_j)\rho_{pj}-\kappa(\delta_{p,M+1}+\delta_{j,M+1}))\rho_{pj}\nonumber\\
&-&i\Gamma(\delta_{p,M+1}\sum_{l=1}^M\rho_{lj}+(1-\delta_{p,M+1}) \rho_{M+1 j}) \nonumber\\
&+&i\Gamma(\delta_{j,M+1}\sum_{l=1}^M\rho_{pl}+(1-\delta_{j,M+1}) \rho_{p M+1}) \nonumber\\
&+&\left({\mathcal O}^{0*}_{pj}-{\mathcal O}^{0*}_{jj}+{\mathcal O}^0_{pj}-{\mathcal O}^0_{pp}\right) \rho_{pj}\nonumber\\
&+&i\left({\mathcal O}^{1*}_{pp}-{\mathcal O}^{1*}_{jp}+{\mathcal O}^{1}_{pj}-{\mathcal O}^{1}_{jj}\right)\sum_{l=1}^MJ_{pl}\rho_{lj}\nonumber\\
&-&i\sum^M_{l=1}\left({\mathcal O}^{1*}_{pl}-{\mathcal O}^{1*}_{jl}+{\mathcal O}^{1}_{lj}-{\mathcal O}^{1}_{lp}\right)J_{lp}\rho_{lj},
\label{master2}
\eea
where $\rho_{pj}=\langle p|\rho_S|j\rangle$, and in this case $\{p,j\}$ run from $1$ to $M+1$. Note however, that the last terms correspond to dephasing, and therefore only appear for $\{p,j\}\neq M+1$. 

\section{Checking the stochastic equations: A dimer}


Before analysing the full LH1-RC system, we study the evolution of two molecules according to the Hamiltonian (\ref{space}). We consider the evolution of the population of the molecule one, $P_1(t)=\langle\sigma_1^\dagger(t)\sigma_1(t)\rangle$, by using the master equation (\ref{master2}), as well as the SSE (\ref{stoch}) for different number of trajectories $NM$. As shown in Fig. (\ref{Twomolec}) when the number of trajectories $NM$ is large enough both results are coincident. This illustrates the convergence of the Montecarlo method, as well as the fact that both schemes (master equation and SSEs) are basically equivalent. In other words, SSEs provide a good alternative to master equations to describe the dynamics of the system.

\begin{figure}[ht]
\centerline{\includegraphics[width=0.45\textwidth]{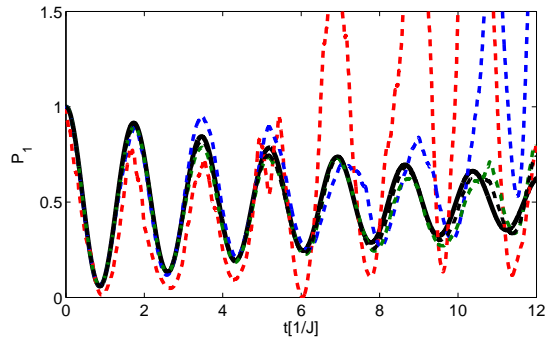}}
\caption{Comparison between master and SSE. The figure displays the time evolution of the population in the molecule $1$ of a dimer, $P_1=\langle \sigma_1^\dagger (t)\sigma_1(t)\rangle$. Solid black line corresponds to the solution for the master equation, while red, blue, green and black dashed lines correspond to $NM=1,100,500,1000$ trajectories of the SSE respectively. For both molecules, the coupling constant $g=0.3$, the energies are $\omega_1=\omega_2=0$, and the decaying rate of the correlation function is $\gamma=10$. The couplings are chosen as $J_{11}=1.5$, $J_{22}=1$ and $J_{12}=1.8$. There is no reaction center in the dimer, so that $\Gamma=0$. \label{Twomolec}}
\end{figure}


\section{Dephasing in the LH1-RC complex}

As noted above, when including dephasing in the description of the transport process, a new mixing mechanism between different momentum components appears, what gives rise to an improvement of the transport efficiency with respect to the closed system case. 
In order to study qualitatively the effect of dephasing, we will consider that all the molecules have the same energy. This approximation is only valid when the molecular energy spread is small as compared to its mean value. However, it will allow us to describe the effects of the dephasing in the system dynamics in an separated way with respect to the effect of energy spread analysed in the former section.

As we have seen in former sections that when considering a closed system with zero dephasing, and uniform energies and couplings to the RC, the total population of non-symmetric momentum components $q\neq 0$, $P_{NS}(t)=\sum_{q\neq 0} \langle\sigma^+_{q}(t)\sigma_{q}(t)\rangle$ is a constant quantity through the evolution. This can be observed in the upper Fig. (\ref{deph}). On the contrary, when dephasing occurs, this quantity decreases severely, as can be seen in the curve corresponding to $NM=500$ trajectories in the upper Fig. (\ref{deph}). 
This in turn is reflected as an increase of the population transfer $P_T$ in an almost specular way (curve $NM=500$ in the lower Fig. (\ref{deph})).  Along this section we will chose a smoothly decaying hopping, $J_{pj}=\frac{1}{(p-j)d_0}$, with $d_0=1/5$, that phenomenology describes the decaying of the hopping with the molecular distance. The coupling from antenna molecules to the reaction center is chosen as  $\Gamma=0.5$. As before, all quantities are in units of $J$.

A brief comment concerning the sampling of SSEs is here in order. In both Figs. (\ref{deph}), it can be seen that for a number of trajectories $NM>250$ the curves converge into a fixed value. This shows that for this number of trajectories, the Montecarlo integrals have been correctly sampled, obtaining the final values for the relevant quantities.

\begin{figure}[ht]
\centerline{\includegraphics[width=0.45\textwidth]{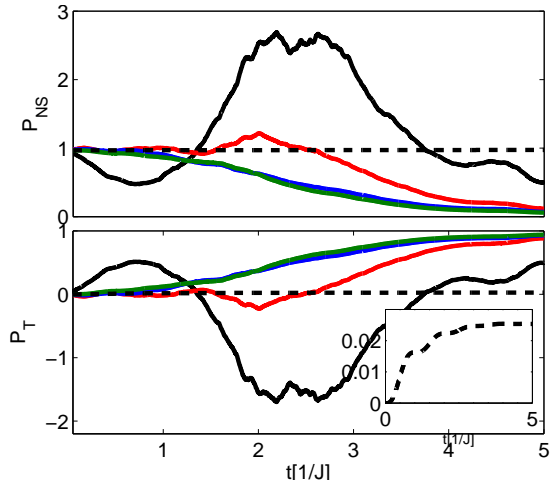}}
\caption{Upper and lower figures represent respectively the evolution of the sum of the non-symmetric momentum components $P_{NS}$ and the transmission probability $P_T$ for different number of stochastic trajectories $NM$. 
The inset of the later shows a detail of $P_T$ for the closed system ($g=0$) at short times. In both figures, solid black, red, blue and green correspond to $NM=10,50,250,500$ respectively, considering a Markovian situation, with $\Gamma=100$, and a coupling constant $g=0.4$. The dashed black corresponds to $g=0$.  \label{deph}}
\end{figure}

We now analyze the total energy absorbed in the reaction center at $P_T(t)$ at $t=5$ for different decaying rates of the correlation function, $\gamma$ and for different couplings $g$. The time $t=5$ is chosen in such a way that the system has already reached its steady state. In Fig. (\ref{markno}) we analyze two situations: a large environment decay rate $\gamma=100$, such that for every coupling $g$ displayed in the figure, the system is in the Markov regime, with $\gamma\gg \Gamma_{\textmd deph}$ ($\Gamma_{\textmd deph}$ being the system dephasing rate); and a situation in which non-Markovian effects begin to appear, with $\gamma=10$. In this particular system, the results shown in Fig. (\ref{markno}) suggest that when dephasing is produced by a non-Markovian reservoir, the transport efficiency is smaller than in the Markov case. The reason is that non-Markovian interactions preserve the coherence of the system longer, so that the environment is less effective in the task of taking the system out of the invariant state subspace, here spanned by the set of $|q\rangle$ with $q\neq 0$. In other words, a non-Markovian reservoir is less efficient on mixing different momentum components than a Markovian reservoir. Although this behavior might be strongly model dependent (in the sense that different choices of the system parameters might give rise to different results), a similar effect is observed in \cite{pleniohuelga09_2} for the FMO complex. 

We should emphasize here that the post-Markov approximation used in this work assumes that the electron hopping time scale is much larger than the correlation time of the environment. This condition might not be fulfilled in realistic photosynthetic complexes, and therefore a more complete analysis is needed to describe the full non-Markovian behavior of the system. Some recent efforts in this direction can be found in \cite{ishizaki09,ishizaki08,ishizaki09_2,prior10}.


\begin{figure}[ht]
\centerline{\includegraphics[width=0.45\textwidth]{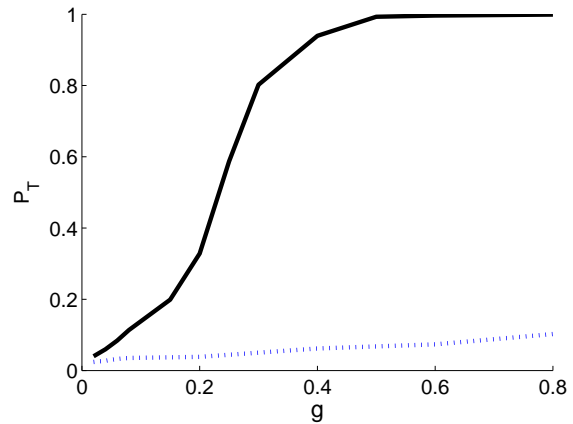}}
\caption{Total energy absorbed in the reaction center $P_T(t)$ at $t=5$ for a Markovian case, $\gamma=100$ (solid black line), and a non-Markovian case, $\gamma=10$ (dashed blue). Here, we only reach intermediate values of the coupling parameter $g$, where the Zeno regime described in \cite{pleniohuelga09_1,rebentrost09} is not observed yet. We consider that this intermediate regime is the most intersting one, since it is the one where the transport efficiency is improved. \label{markno}}
\end{figure}

\section{Conclusions}

By considering a momentum space representation, we have analyzed in a systematic way several mechanisms that may affect the transport dynamics within the LH1-RC complex. To this order, we have considered a simple situation in which all antenna molecules are uniformly coupled to the reaction center, showing that, if additionally the molecular energies are uniform and there is no dephasing, the final transmission probability corresponds to the initial population in the momentum $q=0$, $P_T(t=\infty)=\langle\sigma_{q=0}^\dagger (0)\sigma_{q=0}(0)\rangle$. In other words, only the zero momentum component of the initial state is transmitted to the reaction center. On the other hand, we have shown how energy non-uniformities and dephasing produce energy transfer (or "mixing") from other momentum $q\neq 0$ into the reaction center or sink, increasing the transmission efficiency. 

In order to analyze the effects of the dephasing in the system transport properties, SSEs are used to describe the system evolution. This scheme provides some computational advantages with respect to master equation techniques, inasmuch as it evolves vectors instead of matrices. In addition to that, it allows to explore the system within the so-called post-Markov approximation, that is valid for environment correlation times shorter than the system electron hopping. When comparing the transmitted energy at $t=5[1/J]$ both for Markovian and non-Markovian interactions, it is found that at least in the present system, there is less transmission in the last case. This results are in accordance to what is found in \cite{pleniohuelga09_2}. Indeed, while dephasing destroys undesired coherences in the system, such that it can evolve out of the invariant subspaces and transmit more energy to the reaction center, non-Markovian interactions provide a mechanism to rebuild the coherences. Hence, non-Markovian effects may diminish the positive effects of the environment that exists for intermediate couplings.

We note that even when the coupling of antenna molecules to the reaction center (or reaction centers) is non-uniform, we can still use the momentum representation to determine which momentum values are most efficiently transfered into the reaction center, and which of them are not directly connected to it, and will need some momentum mixing mechanisms (like energy non-uniformities or dephasing) to transfer their energy into the sink.


We thank D. Alonso, J.I. Cirac, A. Ekert, S. Huelga and M.B. Plenio for support and encouragement, and D. Alonso, M.C. Banuls, A. Chin, A. Ekert, G. Giedke,  S. Huelga, M. Roncaglia, C. Navarrete-Benlloch and V. Vedral for interesting and useful dicussions.


\section{Appendix A: Deriving the evolution equation of the reduced propagator}
\label{apenA}
Let us give more details about the derivation of the evolution equation for the reduced propagator (\ref{chapdos13}). From the definition of the reduced propagator, its derivative is given by
\begin{equation}
\frac{\partial G(z^*_1 z_0|t0)}{\partial t}=
\bigg\langle z_1 \bigg| \frac{\partial {\mathcal U}(t,0)}{\partial t}\bigg| z_0 \bigg\rangle.
\label{chapdos7}
\end{equation}
The evolution operator satisfies the Schr\"odinger equation in the partial interaction picture
\bea
\frac{\partial {\mathcal U}(t,0)}{\partial t}&=&\bigg(-i H_S -i \sum_{j \lambda} g_{j\lambda} (L_j^\dagger a_\lambda e^{-i \omega_\lambda t_j}\nonumber\\
&&+ L_j a_\lambda^\dagger e^{i \omega_\lambda t})\bigg){\mathcal U}(t,0).
\label{chapdos9}
\eea

When inserted in (\ref{chapdos7}) this equation leads to
\begin{eqnarray}
\frac{\partial G(z^*_1 z_0|t0)}{\partial t}=&&
\bigg(
-i H_S-i \sum_{j\lambda}L_j g_{j,\lambda} e^{i \omega_\lambda t} z^*_{1,\lambda}
\bigg)\cr &\times&G(z^*_1 z_0|t0) 
-i L_j^\dagger \sum_\lambda g_{j,\lambda} e^{-i \omega_\lambda t}\cr&\times&
\langle z |{\mathcal U}(t,0) a_\lambda(t,0) | z_0 \rangle,
\label{chapdos10}
\end{eqnarray}
where we have used the property $\langle z_1|a^\dagger_\lambda=\langle z_1|z^*_{1,\lambda}$.
In the last term, we have replaced the matrix element 
$\langle z |a_\lambda {\mathcal U}(t,0) | z_0 \rangle$ by $\langle z |{\mathcal U}(t,0) a_\lambda(t,0) | z_0 \rangle$, with $a_\lambda(t,0)={{\cal U}}^{-1}(t,0) a_\lambda {\mathcal U}(t,0)$. This term can be further simplified by integrating the Heisenberg equations of motion for $a_\lambda(t,0)$,
\begin{eqnarray}
\frac{d}{dt }a_\lambda (t ,0 )=
-i \sum_j g_{j,\lambda} e^{-i\omega_\lambda t}L_j(t ,0),
\end{eqnarray}
with $H_0=H_S+H_B$, that leads to
\begin{equation}
a_\lambda(t,0)=a_\lambda(0,0)-i \sum_j g_{j,\lambda} \int_{0}^{t} d\tau L_j(\tau,0) e^{i \omega_\lambda \tau},
\label{chapdos11}
\end{equation}
with
\begin{equation}
L_j(t',t)=e^{iH_Bt}e^{-iH(t-t')}L_j e^{iH(t-t')}e^{-iH_Bt'}.
\label{chapdos12}
\end{equation}
Plugging equation (\ref{chapdos11}) in (\ref{chapdos10}), we get the equation (\ref{chapdos13}) in the paper.










\section{Appendix B: Series expansion of ${\mathcal T}$}
\label{AppendB}

Let us now derive the terms ${\mathcal T}_n$ with $n=0,1,2$ appearing in the evolution equation (\ref{chapdos19}). From the Heisenberg equation for $L_j(\tau-t)$, we can write
\begin{eqnarray}
\frac{dL_j(\tau-t )}{d\tau }=i e^{i H_B t}e^{i H_{tot} (\tau-t)}[H_{tot},L_j]e^{-i H_{tot} (\tau-t)}e^{-i H_B t}
\label{chapdos341}
\end{eqnarray}
Considering the a system Hamiltonian $H_S$ given by (\ref{space}), we find that
\begin{eqnarray}
\frac{dL_j(\tau-t )}{d\tau }&=&-i \bigg(\sum_l J_{jl}\sigma^+_j(\tau-t)\sigma_l(\tau-t)\cr
&-&\sum_l J_{lj}\sigma_l^+(\tau-t)\sigma^-_j(\tau-t)\bigg).
\label{chapdos342}
\end{eqnarray}
This quantity, evaluated in $\tau=t$ as required in (\ref{eleideriva}) can be written as
\begin{eqnarray}
\left.\frac{dL_j(\tau-t )}{d\tau }\right|_{\tau =t}&=&-i\bigg(\sum_l J_{jl}\sigma^+_j\sigma_l^-
-\sum_l J_{lj}\sigma_l^+\sigma^-_j\bigg),
\label{chapdos343}
\end{eqnarray}
where we consider the notation $\sigma^-_j(0)=\sigma^-_j$. In the same way, the second derivative can be evaluated and leads to
\begin{eqnarray}
\frac{d^2L_j(\tau-t )}{d\tau^2 }&=&-i\sum_l J_{lj}\bigg(\left[(\omega_l-\omega_j)+({\mathcal B}^t_l-{\mathcal B}_j^t)\right]{\mathcal A}_{lj}\cr
&+&\sum_p J_{pl}{\mathcal A}_{pj}+\sum_p J_{pj}{\mathcal A}_{pl}\bigg)
\label{chapdos36}
\end{eqnarray}
where we have defined ${\mathcal A}_{pj}=\sigma_p^+\sigma^-_j-\sigma_j^+\sigma^-_p$, and ${\mathcal B}_j^t=\sum_\lambda g_{j\lambda}(a_\lambda e^{-i\omega_\lambda t}+a_\lambda^\dagger e^{i\omega_\lambda t})$.

We now proceed to calculate the expressions for ${\mathcal T}_n$ with $n=0,1,2$, as defined in According the last results and equation (\ref{series}), the zero order ${\mathcal T}_n$  is simply
\begin{eqnarray}
{\mathcal T}_0=\sum_{p,j=1}^M\int_0^t d\tau \alpha_{pj}(t-\tau)L_p^\dagger L_j G(z^*_1 z_0|t0).
\label{series0A}
\end{eqnarray}
where $G(z^*_1 z_0|t0)=\langle z | {\mathcal U}(t,0)|z_0 \rangle$.
The first order can be obtained by just inserting (\ref{chapdos343}) in (\ref{series}), so that
\begin{eqnarray}
{\mathcal T}_1&=&
i\sum_{p,j=1}^M\int_0^t d\tau \tau{\mathcal R}_{pj}(\tau) G(z^*_1 z_0|t0)
\label{series1A}
\end{eqnarray}
where in our case
\bea
{\mathcal R}_{pj}(\tau)=\bigg(J_{pj}\alpha_{pp}(\tau)\sigma_j^+\sigma_p
-J_{pj}\alpha_{pj}(\tau)\sigma^+_p\sigma_j\bigg)
\eea
The second line has been obtained after considering $L_j^\dagger=\sigma^+_j\sigma_j$ and a single excitation in the system.
The second order has the following form
\begin{eqnarray}
{\mathcal T}_2&=&\sum_{pl}\int_0^t d\tau \tau^2 {\mathcal R}_{pl}(\tau)
\bigg\{{\mathcal C}_{lp}(z^*_1z_0)+(\omega_l-\omega_p)G(z^*_1 z_0|t0) \bigg\}\cr
&+&\int_0^t d\tau \tau^2 ({\mathcal P}_{pkl}(\tau)+{\mathcal Q}_{pkl}(\tau))G(z^*_1 z_0|t0),
\label{series2A}
\end{eqnarray}
where
\bea
{\mathcal C}_{lp}(z^*_1z_0)=\langle z_1|({\mathcal B}^t_l-{\mathcal B}^t_p){\mathcal U}(t,0)|z_0\rangle
\eea
and
\begin{eqnarray}
{\mathcal P}_{pkl}(\tau)&=&J_{pl}J_{kl}\left\{\alpha_{pk}(\tau)\sigma^+_k\sigma^-_p
-\alpha_{pp}(\tau)\sigma_p^+\sigma^-_k\right\};\cr
{\mathcal Q}_{pkl}(\tau)&=&J_{pl}J_{kp}\left\{\alpha_{pk}(\tau)\sigma^+_k\sigma^-_l
-\alpha_{pl}(\tau)\sigma_l^+\sigma^-_k\right\}
\end{eqnarray}
All the terms in the former equation are linear functions of the reduced propagator $G(z^*_1 z_0|t0)$, except for those containing the environment operators ${\mathcal B}_j$, whose matrix elements should be correctly evaluated. In particular, we need to compute ${\mathcal C}_{pl}(z^*_1z_0)$,
\begin{eqnarray}
{\mathcal C}_{pl}(z^*_1z_0)&=&\sum_\lambda (g_{l,\lambda}-g_{p,\lambda})\left(z_{1,\lambda}^*e^{i\omega_\lambda t}G(z^*_1 z_0|t0)\right.\nonumber\\
&+&\left.e^{-i\omega_\lambda t}\langle z_1|{\mathcal U}(t,0)a_\lambda(t,0)| z_0\rangle \right).
\end{eqnarray}
Replacing equation (\ref{chapdos11}) in the matrix element $\langle z|{\mathcal U}(t,t0)a_\lambda(t,0)| z_0\rangle$, we get
\begin{eqnarray}
{\mathcal C}_{pl}(z^*_1z_0)&=&\sum_\lambda (g_{l,\lambda}-g_{p\lambda})\bigg((z_{1,\lambda}^*e^{i\omega_\lambda t}+z_{0\lambda} e^{-i\omega_\lambda t})G(z^*_1 z_0|t0)\cr&-&i \sum_l e^{-i\omega_\lambda t}\langle z_1|{\mathcal U}(t,0)L_l(\tau,0)|z_0\rangle \bigg).
\label{series4}
\end{eqnarray}
Again, we may consider that $\langle z_1|{\mathcal U}(t,0)L_l(\tau,0)|z_0\rangle=\langle z_1|L_l(\tau,t){\mathcal U}(t,0)|z_0\rangle$, and a series expansion of the operator $L_l(\tau,t)$ in terms of $(\tau-t)$. Note that the term (\ref{series4}) is a component of ${\mathcal T}_2$, a second order term in the expansion. Hence, we just need to consider $L_l(\tau,t)\sim L_l +{\mathcal O}(n)$  in (\ref{series2A}). The higher orders in $n$ would give rise contributions of orders $n>2$, which here are discarded.

With this considerations, and replacing (\ref{series4}) in (\ref{series2A}) we get the following expression for the second order term of the expansion,
\begin{eqnarray}
{\mathcal T}_2&=&\bigg(\sum_{pl}\int_0^t d\tau \tau^2{\mathcal R}_{pl}(\tau)\bigg\{\sum_\lambda (g_{l,\lambda}-g_{p,\lambda})\eta_\lambda+(\omega_l-\omega_p) \bigg\}\cr
&+&\sum_{plk}\int_0^t d\tau \tau^2({\mathcal P}_{pkl}(\tau)+{\mathcal Q}_{pkl}(\tau))\bigg)G(z^*_1 z_0|t0)
\label{series5}
\eea
where
\bea
\eta_\lambda=\left(z_{1,\lambda}^*e^{i\omega_\lambda t}+z_{0\lambda} e^{-i\omega_\lambda t}+ i \sum_l L_l e^{-i\omega_\lambda t}\right).
\eea


\section{Appendix C: Validity conditions of the post-Markov approximation}

In this paper we have analyzed the system dynamics by considering the so-called post-Markov approximation. Let us now derive the validity conditions for this approximation, and compare this conditions to the ones required in the weak-coupling approximation, and in the so-called F\"orster theory.

Indeed, the post-Markov approximation is valid only when the series expansion of ${\mathcal T}=\sum_n {\mathcal T}_n$ converges at small $n$. This requires in particular that, for any $j$, the function
\begin{eqnarray}
{\mathcal F}_n(t)=S^n \int_0^t dm m^n \alpha^T_{jj} (m)
\label{condition}
\end{eqnarray}
with $S\sim(J_{jj}, \omega_j-\omega_p)|_{\textmd max}$ is a monotonically decreasing function of $n$. The requirement can be written as
\begin{eqnarray}
{\mathcal F}_n (t)>{\mathcal F}_{n+1}(t)
\label{condition2}
\end{eqnarray}
for any $n$ and any time $t<t_{\textmd max}$, where $t_{\textmd max}$ is the maximum time at which the system is evolved.

Let us now assume the environmental correlation function (\ref{correl}) considered in this paper. With this correlation, the function (\ref{condition}) can be written as
\begin{eqnarray}
{\mathcal F}_n(t)=\left(\frac{S}{\gamma}\right)^n \frac{\Gamma[1+n]-\Gamma[1+n,t\gamma]}{\gamma}
\label{condition3}
\end{eqnarray}
where $\Gamma[z]=\int_0^\infty dt e^{-t}t^{z-1}$ is the Euler gamma function, and $\Gamma[a,z]=\int_z^\infty dt e^{-t}t^{z-1}$ is the incomplete gamma function.
Clearly, the function (\ref{condition3}) is a monotonically decreasing function only if the ratio $\frac{S}{\gamma}\ll1$. In terms of the electronic hopping $J$, the condition can be written as
\bea
\frac{1}{J}\gg\frac{1}{\gamma}.
\label{condition4}
\eea
In other words, it is required that the relaxation time of the environment is much shorter than the electron hopping. A further requirement is to chose $\gamma$ small enough, so that for the maximum time $t_{\textmd max}$, the increasing function in $n$, $\Gamma[1+n]-\Gamma[1+n,t\gamma]$, is compensated by the decreasing function $\left(\frac{S}{\gamma}\right)^n$.
From this discussion, it is clear that in principle within the post-Markov approximation the coupling between system and environment can be chosen arbitrarily. Therefore, it works even when the dephasing time scale of the system is comparable (or smaller) than the hopping time. In addition, for systems with small enough hopping, it may be even valid for cases in which the dephasing time of the system is comparable to the environment correlation time.



The post-Markov approximation is an alternative to the weak coupling approximation and the so-called F{\"o}rster theory, since it is valid in different parameter regimes. In the weak coupling approximation, the coupling strength between the system and the environment is considered to be small with respect to the magnitude of system and environment Hamiltonians ($g\ll S,B$, where $S=||H_S||$ and $B=||H_B||$). Furthermore, along with the weak coupling approximation it is often assumed that the environment correlation time $\tau_c\sim 1/\gamma$ is much smaller than the evolution time scale of the system undergoing dephasing $1/\gamma_{\textmd deph}$, i.e.
\bea
\frac{1}{\Gamma_{\textmd deph}}\gg \frac{1}{\gamma}
\eea
where $\Gamma_S\sim g$ in terms of our parameters, so that the condition can be written as $\gamma\gg g$. In this approximation the coupling $J$ and the energy detunings are arbitrary.

In the F{\"o}rster theory \cite{ishizaki09}, a strong electronic coupling is considered, such that that electronic hopping occurs in a time scale much smaller than the evolution time scale of the system undergoing dephasing,
\bea
\frac{1}{\Gamma_{\textmd deph}}\gg\frac{1}{J},
\eea
or $J\gg\lambda$. The correlation time of the environment, although somehow related with $\Gamma_S$ is in principle irrelevant.
\section*{References}
\bibliography{Biobib}
\bibliographystyle{unsrt}

\end{document}